\documentclass{article}
\usepackage{blindtext}
\usepackage[a4paper, total={6in, 8in}]{geometry}

\usepackage{graphicx}       

\usepackage{physics}
\usepackage{lineno}
\usepackage{siunitx}
\usepackage[free-standing-units=true]{siunitx}
\usepackage[hidelinks]{hyperref}
\usepackage{cite}
\usepackage{bm}
\usepackage{bbold}
\usepackage{authblk}

\title{\textbf{Reconfigurable unitary transformations of optical beam arrays}}

\author[1,2,*]{ Aldo C. Martinez-Becerril }
\author[1]{Siwei Luo}
\author[1]{ Liu Li}
\author[1]{ Jordan Pag\'e }
\author[3]{ \\ Lambert  Giner }
\author[4]{ Raphael A.  Abrahao}
\author[1]{ Jeff S. Lundeen }

\affil[1]{Department of Physics and Nexus for Quantum Technologies, University of Ottawa,
   25 Templeton St, Ottawa, ON, K1N 6N5 Canada }
\affil[2]{The Edward S. Rogers Department of Electrical and Computer Engineering, University of Toronto, 10 King’s College Road, Toronto, ON M5S 3G4, Canada}
\affil[3]{ D\'epartement de Physique et d’Astronomie, Universit\'e de Moncton, 18 Ave. Antonine-Maillet, Moncton, NB E1A 3E9, Canada }
\affil[4]{Brookhaven National Laboratory, Upton, New York 11973, USA}
\affil[*]{aldo.martinezbecerril@utoronto.ca}

\date{}

\begin{document}
\maketitle

\begin{abstract}
Spatial transformations of light are ubiquitous in optics, with examples ranging from simple imaging with a lens to quantum and classical information processing in waveguide meshes.  Multi-plane light converter (MPLC) systems have emerged as a platform that promises completely general spatial transformations, i.e., a universal unitary. However until now, MPLC systems have demonstrated transformations that are far from general, e.g., converting from a Gaussian to Laguerre-Gauss mode. Here, we demonstrate the promise of an MLPC, the ability to impose an arbitrary unitary transformation that can be reconfigured dynamically. Specifically, we consider transformations on superpositions of parallel free-space beams arranged in an array, which is a common information encoding in photonics. We experimentally test the full gamut of unitary transformations for a system of two parallel beams and make a map of their fidelity. We obtain an average transformation fidelity of $0{.}85 \pm 0{.}03$. This high-fidelity suggests MPLCs are a useful tool implementing the unitary transformations that comprise quantum and  classical information processing.
\end{abstract}

\section{Introduction}\label{section:Intro}
The transformation of optical spatial modes in a reconfigurable and reliable way would have a wide range of applications, ranging from fundamental studies in optics to applications in telecommunications, classical computing, and quantum information processing. Most passive optical information processing tasks are unitary transformations $\bm{U}$ (i.e., $\bm{U}\bm{U^{\dagger}} = \bm{U^{\dagger}}\bm{U} = \bm{1}$) since they preserve information in the field. A device that creates a reconfigurable unitary could implement matrix multiplication \cite{WOS:000750832600001}, optical quantum gates \cite{Brandt:20, PhysRevApplied.18.014063} and quantum random walks \cite{Jiao:21,esposito2022quantum,schreiber20122d}. Furthermore, adding a non-linear optical element to such a device would allow one to obtain an optical neural network \cite{shen2017deep,zuo2019all,feldmann2019all, huang2021silicon, Jha:20}. In this work, we experimentally implement arbitrary unitary transformations in the space of a two-beam array (two parallel optical beams)  using a platform known as multi-plane light converter (MPLC) \cite{Morizur:10, Labroille:14, armstrong2012programmable}. 

An N-dimensional unitary $\bm{U}$ possess $N^2$ independent real parameters that need to be specified to fully determine $\bm{U}.$ To physically implement a unitary on spatial modes such as optical beams, one typically needs two elements: a mode coupler and the ability to impart phases to each mode. Two approaches can be distinguished, one that uses two-mode beamsplitters as mode couplers, and another that uses an optical Fourier transform or diffraction to couple modes. 

In optics, the first approach was introduced in a seminal paper \cite{PhysRevLett.73.58}, the Reck scheme, which demonstrated a deterministic algorithm to decompose a general unitary of dimension $N$ into phase shifters and beamsplitters. The result is a `mesh' of beamsplitters that composes a large interferometer with $N^2$ arms. Each arm length needs to be mechanically stabilized to much less than a wavelength. All together, a stable mesh is a monumental challenge, particularly for modes travelling in free-space \cite{PhysRevLett.91.187903}. Instead, almost all demonstrations have used integrated optic waveguide meshes, with reconfigurability provided by thermal phase-shifters. With the first such a device, unitary transformations of dimension $N = 6$ were used for implementing quantum gates, boson sampling, and random unitary transformations \cite{doi:10.1126/science.aab3642}. This approach has been extended to create reconfigurable circuits with $N = 8 $ \cite{Taballione:19}, $N=12$ \cite{taballione2021universal} and $N=20$ \cite{Taballione2023modeuniversal} waveguides. Additionally, an alternative to the Reck scheme was introduced by Clements \textit{et al.} \cite{Clements_Optica2016} that minimizes the depth of the mesh. The beamsplitter mesh approach has proven to be enormously influential in both quantum and classical information processing.

We now turn to the diffraction approach. An MPLC is composed of layers alternating between a phase-mask plane and a diffraction layer. The latter is either created by free-space propagation or an optical Fourier transform. The phase-mask planes have been typically implemented using a spatial light modulator (SLM). The MPLC was conceived and demonstrated in Refs.~\cite{Morizur:10, Labroille:14, armstrong2012programmable}. Initially, Refs.~\cite{Morizur:10,Labroille:14} focused on the use of MPLCs to perform mode multiplexing, and the phase-mask profiles were obtained by numerical optimization. In Refs.~\cite{Design_WaveFront_8346129,fontaine2019laguerre} a particular numerical optimization, `wavefront matching', was adapted to MPLCs. They converted each beam in a two-dimensional grid of beams to a different Laguerre-Gauss optical mode; achieving the impressive result of converting $210$ optical modes using seven phase-mask planes. In the recent years, there has been an increasing number of works that use MPLCs to perform different tasks such as diffractive networks \cite{doi:10.1126/science.aat8084}, high-dimensional quantum gates using orbital angular momentum \cite{Brandt:20,HighD2PhotonInterference}, processing entanglement in high-dimensions \cite{ProcessingEntangledPhotonsInHighD}, spatial mode multiplexing \cite{HighDSpatialModeSortingOptCircuitDesign,Rouviere:24}, control of multiple degrees of freedom of light \cite{mounaix2020time,cruz2022synthesis}, and  sorting states of light \cite{SortingOverlappingStatesOfLight}. Other works are looking at different platforms like dielectric slabs \cite{Larocque:21}. The MPLC has drawn much interest for both applications and fundamental science.

An MPLC presents a series of advantages over the beamsplitter approach. To mention a few: (i) in an MPLC one has more phase control channels, as there are of the order of $10^6$ pixels on a single SLM. In the beamsplitter approach, the number of phase controls is, in practice, limited to order of $10^2$ \cite{taballione2021universal}; (ii) An MPLC can be fully implemented in free-space, thus it does not suffer from insertion losses to chip and waveguides;  (iii) Rather than being a microfabricated device, an MPLC is composed of common optical elements, thereby making it implementable in a larger number of laboratories. The simple design is also simpler to scale to higher dimension states.

One of the main drawbacks of the MPLC compared to the beamsplitter mesh is the lack of a deterministic analytical algorithm for finding the phase-mask profiles. Consequently, in an MPLC system, the minimum number of phase-mask planes needed to create a given unitary of dimension $N$ is unknown. That said, if there are $N$ pixels in each phase-mask, there must be at least $N$ planes so that one can set all the $N^2$ free parameters of the unitary. This sets a bound for the minimum number of required phase-mask planes. The seminal work \cite{borevich1981subgroups} provided an existence proof that an MPLC could implement a general unitary transformation. More recent work \cite{LopezPastor:21} gave an analytical method to implement any unitary but it requires $6N$ phase-mask planes, far from the minimum, albeit all but $N$ were fixed phase-mask profiles. At present, there is no analytical MPLC design for an N-plane $N^2$ dimensional unitary.

Compounding the lack of a analytical design is that, unlike the discrete space theory mentioned above, a real MPLC acts in a continuous position space to perform transformations. Moreover, in practice, diffraction replaces the Fourier transform used in the above mentioned theory. In short, our analytical understanding of the MPLC is far-removed from how it is used in practice. The lack of an analytical algorithm means that the optimal transformation fidelity and its dependence on the MPLC system parameters are unknown. With so much unknown about MPLC design and performance it is important to characterize it for a wide range of tasks. The origin of this complication is that an MPLC uses a continuous position space to perform the transformations, and diffraction is hard to analytically deal with under the restriction that the MPLC works entirely on a discrete space.

Most previous implementations of MPLC systems have created a single chosen transformation. Moreover, usually that transformation was special in some way, e.g., Laguerre-Gauss modes are propagation eigenstates.  It was unclear how well an MPLC would perform for an arbitrary target unitary. In this work, we experimentally demonstrate an MPLC's versatility and reconfigurability by  implementing a set of transformations that densely samples and fully covers the space of possible unitary transformations. In this way, we show that an MPLC can be a universal unitary.

Like the original Reck scheme, we aim to create unitary transformations of a linear array of parallel beams (a beam array). We consider a beam array because it is a common information encoding in photonics for classical and quantum information processing. Moreover, a \textit{linear} array matches the square grid of pixels in an SLM. Specifically, a linear beam array can be aligned with the first column of the grid so that subsequent phase-mask planes use subsequent SLM columns. Or, more abstractly, if the SLM has $N^2$ pixels, that would be sufficient, in principle, to implement a unitary of dimension $N$. In short, using a beam array on a line nominally makes an effective use of the active area of the SLM, which could be helpful to scaling an MPLC to higher-dimensional transformations. 

We continue in the next section by introducing the MPLC in more detail, formally defining a beam array, and describing the wavefront matching algorithm that we use to find our phase-mask profiles for each unitary in our full set. We then proceed to describe our experimental setup and to characterize its performance using the design phase-mask profiles, thereby creating a map of the transformation fidelity over all unitary transformations. Finally, we summarize our work and point future research directions.  

\section{Multi-plane light converter (MPLC)}
A single phase-mask is insufficient to create a general spatial unitary. To see this consider a general two-mode unitary transformation (Eq.~\eqref{eq:2DUnitary}). Any thin phase-grating will couple more than just these two modes \cite{moharam1978criterion}. For example, a sinusoidal or blazed diffraction grating will diffract light into new orders upon successive applications. This is Raman-Nath diffraction. On the other hand, thick gratings are able to limit coupling to two and only two modes through e.g., Bragg diffraction \cite{moharam1978criterion}, as required by a two-mode unitary. A thick grating can be pictured as a series of different phase-mask planes interleaved with spatial propagation, i.e., diffraction. This arrangement, the MPLC, is exactly what is needed to create a general spatial unitary.

\begin{figure}[h!]
    \centering
    \includegraphics[width = 4 in ]{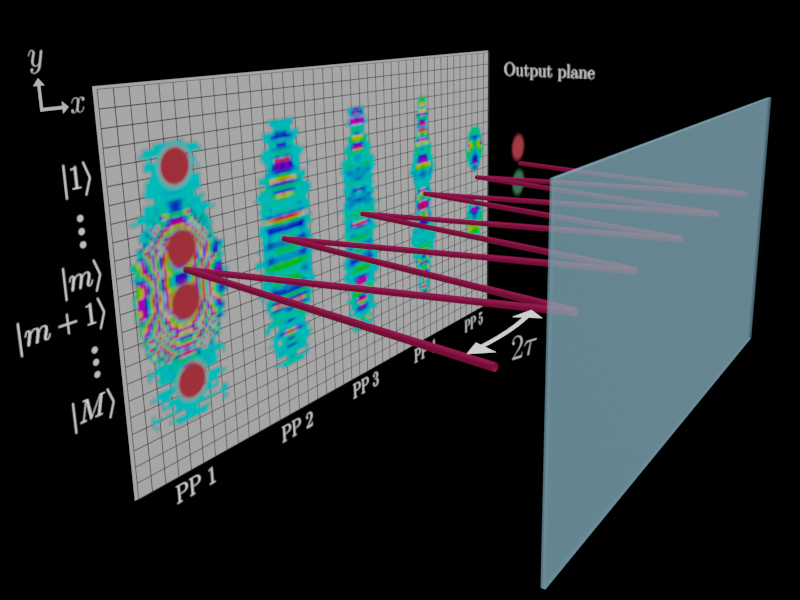}
    \caption{Schematic of a multi-plane light converter (MPLC), composed by a spatial light modulator (SLM) and a flat mirror. These optical components are separated by a distance $L$. In an MPLC, a set of $P$ phase-mask planes (PP) (color indicates phase, red~$=0$ and green~$=\pi$) and diffraction implement a unitary transformation of dimension $N$. A beam array (circles at PP1) is at the input of the MPLC with insertion angle $\tau$. At the output plane after the SLM, we show $\bm{U}\ket{m}$, i.e., the output state after the MPLC when the input is beam $\ket{m}$ of the beam array. }
    \label{fig:MPLC_BeamArray_PMs}
\end{figure}

Applying a phase-mask and diffraction are unitary transformations that we denote by $\bm{\mathcal{P}}$ and $\bm{\mathcal{D}}$, respectively. An MPLC effectively performs a unitary transformation $\bm{U}$ (assuming the numerical aperture of the system is large enough) by concatenating $P$ blocks of $\bm{\mathcal{D}}\bm{\mathcal{P}}$: $\bm{U} =  \bm{\mathcal{D}}_P \bm{\mathcal{P}}_P \ldots \bm{\mathcal{D}}_1 \bm{\mathcal{P}}_1$. A schematic of an MPLC is shown in Fig.~\ref{fig:MPLC_BeamArray_PMs}. A flat mirror and an SLM parallel to each other and $L$ distance apart form the MPLC system. That is, instead of using a separate SLM for each phase-mask plane, we reflect the light back to the same SLM. The light is inserted to the MPLC at an angle $\tau$, with respect to the normal of the SLM, so that each incidence on the SLM is at a different $x$ position, utilizing a different area on the SLM.  In this figure, we also show a linear beam array along $y$ (red circles) as the set of input beams on which the transformation is being performed. 

\subsection{Beam array}
A beam array of dimension $M$ consists of a set of $M$ optical beams $\ket{m}$, where $m$  is used as an index to label each beam $m = 1, \ldots, M $.  We consider a beam array where the centers of each beam are located on a line parallel to the $y$ direction. The beam spacing (center-to-center separation) between two adjacent beams is~$\Delta y$. We focus on the case where the electric field $\psi_m\left(x,y\right) = \braket{x,y}{m}$ of each beam is Gaussian: $\psi_m\left(x,y\right) \propto G\left(x\right)G_m\left(y\right)$, where $G_m\left(y\right) = e^{-\left(y-y_m\right)^2/\omega^2\left(z\right)}$ and $G\left(x\right) = e^{-x^2/\omega^2\left(z\right)}$. Here, the beam size $\omega\left(z\right) $ is the half-width at $1/e^2$ intensity of the beam, which is a function of the propagation distance~$z$. The overlap of two beams exponentially decays with an increasing beam spacing $\Delta y$:  $\braket{m}{m^{\prime}} = \int \psi_m^*\left(x,y\right)\psi_{m^{\prime}}\left(x,y\right) dx dy = e^{- \left(m-m^{\prime}\right)^2\Delta y^2 /\left(2\omega^2\left(z = 0 \right)\right)}$ where~$^*$~indicates complex conjugation. Two beams are nearly orthogonal if  $\Delta y > \sqrt{2}\omega\left(z=0 \right)$, and they become completely orthogonal in the limit $\Delta y \gg \sqrt{2}\omega\left(z=0 \right)$. Thus the beams form an approximate orthonormal basis for the beam array Hilbert space. 
A general state $\ket{\psi}$ in the beam array is a superposition of the $\ket{m}$ beams, i.e., 
\begin{equation} \label{eq:general_beam_array_state}
\ket{\psi} = \sum_{m=1}^M c_m\ket{m},     
\end{equation}
where $c_m = \braket{m}{\psi}$. For a single photon, this state can be understood as the superposition of the amplitudes for the photon to be in each beam. Or, from the viewpoint of a classical coherent field description, each amplitude scales each beam's Gaussian electric field distribution. 

\subsection{Unitary transformations in a beam array}\label{sec:Unitary_2BeamArray}
The  case of a general unitary  transformation $\bm{U}$ on a two-beam array is given by

\begin{equation}
\bm{U}\left(\theta, \phi\right) =  \begin{pmatrix}
    \cos\theta  & ie^{-i\phi}\sin\theta \\
    ie^{i\phi}\sin\theta & \cos\theta  
    \end{pmatrix},
    \label{eq:2DUnitary}
\end{equation}
where $\theta \in [0, \pi/2]$ is the coupling parameter, and $\phi \in [-\pi, \pi)$ is a phase. The other two parameters for a general two-dimensional unitary (see for example Ref.~\cite{murnaghan1962unitary}) are omitted as one of them is a global phase, and the other is a phase that can be factored as a diagonal matrix and can be straightforwardly applied after the unitary by applying a different flat phase shift to each beam. The unitary $\bm{U}\left(\theta, \phi\right)$ is a two-mode beamsplitter transformation. In this context, one identifies $\theta$ with the reflectivity $R$ of the beamsplitter as $|R| = \sin^2\theta$. The case $\theta = \pi/4$ corresponds to a 50:50 beamsplitter. The case $\theta = \pi/4$ and $\phi = -\pi/2$ corresponds to the Hadamard transformation (with the beams swapped) widely used in quantum information \cite{nielsen_chuang_2010}. 

\subsection{Optimization procedure}
Now we describe the general idea behind the optimization procedure to find the required phase-mask profiles for achieving a given unitary. We follow Ref.~\cite{fontaine2019laguerre} and use a wavefront matching algorithm \cite{sakamaki2007new,Design_WaveFront_8346129} to obtain each of the phase-mask profiles in our system. In a general MPLC (as the one shown in Fig.~\ref{fig:MPLC_BeamArray_PMs}), there are $P$ phase-mask planes equally separated by a distance $\Delta z = 2\times L$. The phase distribution at plane $p$ is $\Phi_p\left(x,y\right)$. At the first plane, we have the $M$ input beams ${\ket{m}}$, where $m$ is used as an index to refer to each beam $m = 1, \ldots, M$. The beam array states  $\ket{\psi_{t_m} } $  at the last plane are of the form in Eq.~\eqref{eq:general_beam_array_state} and are determined by the target unitary $\bm{U}_t$ i.e.,  $\ket{\psi_{t_m}} = \bm{U}_t\ket{m}$. We call design states ${\ket{\psi_{d_m} }}$ the theoretical output states of a given MPLC design (i.e., a specific set of designed phase-mask profiles), thus $\ket{ \psi_{d_m} } =  \bm{U}_{d}\ket{m}$, where $\bm{U}_{d} = \bm{\mathcal{D}}_P \bm{\mathcal{P}}_P \ldots \bm{\mathcal{D}}_1\bm{\mathcal{P}}_1$. 

The phase-mask profiles $\Phi_p$ are obtained through an inverse design optimization. The goal of which is to minimize the phase difference between input and target states at every propagation plane by adjusting the phase-mask profiles $\Phi_p$ located at different propagation planes. The phase mismatch of corresponding input-target states at phase-plane one is given by 
\begin{equation}
\int\int \text{arg}\left( \psi_m\left(x, y, \Delta z \right)\psi_{t_m}^{*}\left(x, y, \Delta z \right) \right) dxdy,    
\end{equation}
for each of the $m = 1,\ldots,M$ pairs of input-output states. A superposition of these phase errors is formed each of them being weighted by the overlap of the corresponding input-target states of the MPLC system. The updated phase at phase-mask plane one is given as the phase of such a superposition, this is mathematically written as follows

\begin{equation}\label{eq:PM_Update}
\Delta \Phi_1\left(x,y\right) =  - \arg\left( \sum_{m} \psi_m\left(x,y,\Delta z\right) \psi_{t_m}^{*}\left(x,y,\Delta z\right) e^{ -i\int \text{arg}\left( m\left(x,y,\Delta z\right)\psi_{t_m}^{*}\left(x,y,\Delta z\right) \right) dxdy } \right).    
\end{equation}
A similar procedure is used to update the other phase-mask profiles by propagating the input and target states to the proper plane. For example, to update the $p^\text{th}$ phase-mask profile, one needs to propagate the input beams to this plane as $\bm{\mathcal{P}}_p \ldots \bm{\mathcal{D}}_1 \bm{\mathcal{P}}_1\ket{m}$. A single iteration of the algorithm is conducted by repeating this phase update for all $P$ phase-mask planes. This iteration is then repeated until the figure of merit is reduced below a threshold value or a number of iterations is achieved. The figure of merit is the gate fidelity $F_G$  between the target unitary $\bm{U}_t$ from Eq.~\eqref{eq:2DUnitary} and the one implemented by the design MPLC $\bm{U}_{d}$, 
\begin{equation}\label{eq:GateFidelityDef}
F_G(\bm{U}_t, \bm{U}_{d}) =  \frac{1}{M}\sum_{m = 1}^M \Big{\lvert}\bra{m}\bm{U^{\dagger}}_t\bm{U}_{d}\ket{m}\Big{\rvert}^2. 
\end{equation} 
Thus the gate fidelity can be obtained from the electric field of the target and output design states as 
\begin{eqnarray} \label{eq:GateFide_FromFields}
F_G(\bm{U}_t, \bm{U}_{d}) & = &  \frac{1}{M} \sum_{m = 1}^M \Big{\lvert} \bra{\psi_{t_m}}\ket{\psi_{d_m}} \Big{\rvert}^2  \nonumber \\
& = &  \frac{1}{M} \sum_{m = 1}^M \Big{\lvert} \int\int \psi_{t_m}^{^*}(x,y)\psi_{d_m}(x,y) dxdy \Big{\rvert}^2. 
\end{eqnarray}
For further details of the algorithm we refer the reader to Ref.~\cite{Design_WaveFront_8346129}.

\subsection{Choice of parameters}\label{subsection:Designparameters}
In this subsection, we describe how we selected the geometric parameters of the beam array and the MPLC system. As mentioned in Section~\ref{section:Intro}, the realization of an N-dimensional unitary requires at least $N$ phase-mask planes. We estimated the number of possible reflections on an MPLC system,  details can be found in the \hyperref[sec:supplementarymaterial]{Supplementary material}. For our SLM (dimensions given in Section~\ref{sec:Experiment}) we found there are up to ten reflections that can be achieved with realistic parameters. We used five phase-mask planes for our MPLC. This number is expected to achieve our target transformations from Eq.~\eqref{eq:2DUnitary}. We proceed to perform an optimization of the system using the wavefront matching algorithm as we explain next.

\begin{figure}[htb!]
    \centering
    \includegraphics[width=4 in ]{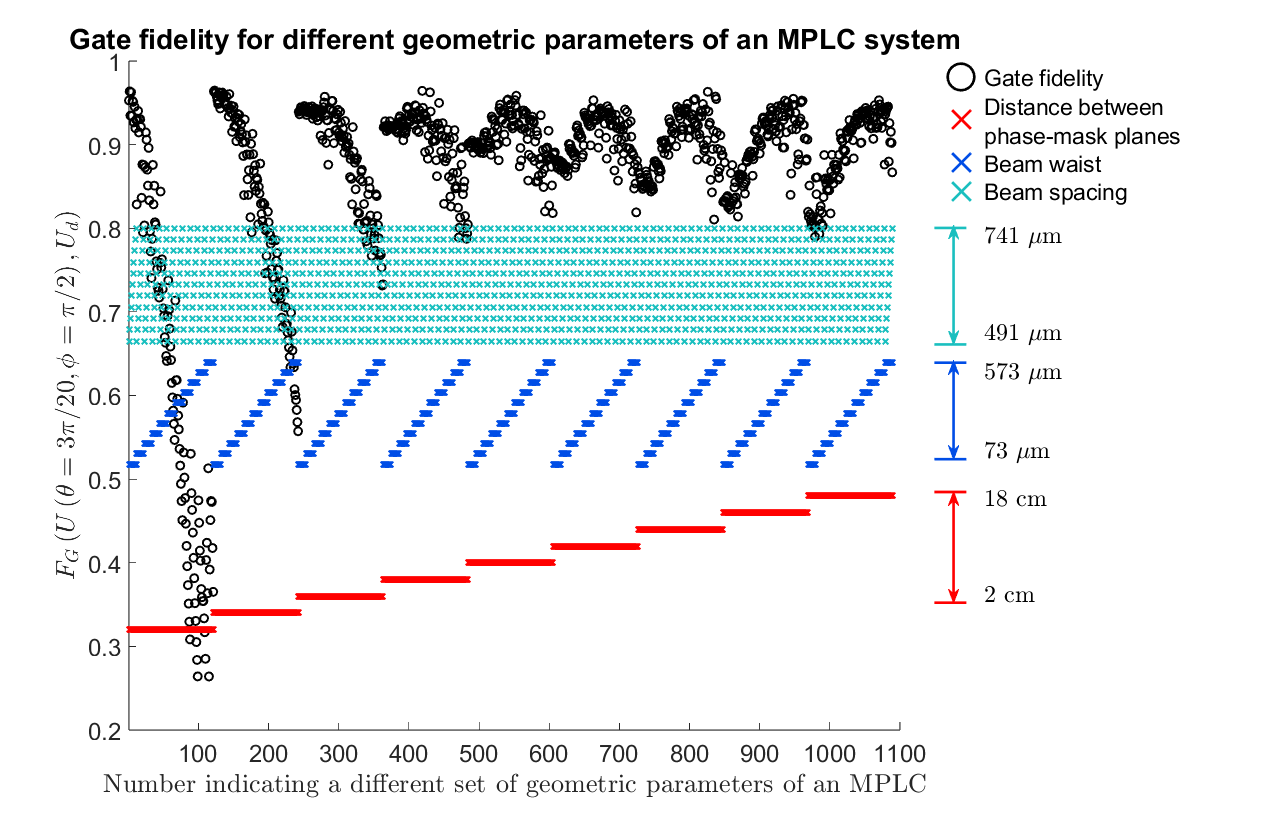}
    \caption{ Theoretical gate fidelity $F_G$ (shown in black markers and obtained using Eq.~\eqref{eq:GateFide_FromFields}) between the target unitary $\bm{U}\left(\theta = 3\pi/20,\phi = \pi/2\right)$ and the one implemented by the design MPLC system with different geometric parameters. The geometric parameters are the following: distance between phase-mask planes (red), beam waist (dark blue), and beam spacing (light blue). The range in which each parameter is sampled is indicated by its maximum and minimum values on the right. For each set of parameters (labeled by an integer index displayed in the x axis) the wavefront matching algorithm was used to find the phase-mask profiles to achieve the target unitary. }
    \label{fig:ChoiceofParameters}
\end{figure}

We used the wavefront matching algorithm \cite{Design_WaveFront_8346129,fontaine2019laguerre} to choose the parameters of an MPLC system that implements the unitary transformations from Eq.~\eqref{eq:2DUnitary}.
Fig.~\ref{fig:ChoiceofParameters} shows the theoretical gate fidelity for the target unitary $\bm{U}\left(\theta = 3\pi/20, \phi = \pi/2\right)$ and the one implemented by different MPLC systems (the different MPLC geometric parameters are given in the caption of Fig.~\ref{fig:ChoiceofParameters}). We found the MPLC achieves a theoretical gate fidelity higher than $ 0.9$ for a wide range of parameters. On the other hand, a spacing between phase-mask planes smaller than  $\SI{5}{\centi\metre}$ combined with a beam waist greater than $\SI{500}{\mu\metre}$  leads to a gate fidelity lower than $0.6$. We selected a spacing between phase-masks of $\SI{10}{\centi\metre}$,  for which the MPLC achieves a gate fidelity of $0.9$ with variations of $5\%$ across the sampled values of beam spacing and beam waist, which suggests the system is robust to experimental imperfections. With the chosen parameters, we verified the MPLC system achieved a similar performance when changing the target unitary. 

We now list the final design parameters. The wavelength is $\SI{637}{\nano\metre}$ and the SLM's pixel size is $\SI{20}{\micro\metre}$. We are using $P = 5$ phase-mask planes and a plane spacing of $2L = \SI{10}{cm}$. Adjacent input beams have a center-to-center separation of $\Delta y=\SI{704}{\micro\metre}$. Each  beam has a waist (half width at $1/e^2$ intensity) of $\SI{161.5}{\micro\metre}$ located $\SI{2.08}{\cm}$ before the first phase-mask plane. The overlap between both beams equals $\SI{7.5 e-5}{}$, thus they are nearly orthogonal. We apply a linear phase $e^{iy\pi/\left(18 \times \SI{20}{\micro\metre}\right)}$ on one of the beams to account for the beam tilt experimentally observed in our setup. The output beam array is the same as the input one with the beam waist and beam spacing demagnified by a factor of four. We observed that this demagnification helps the MPLC to compensate for the diffraction of the beams and reduces losses due to diffraction out of the light path.

\subsection{Optimization results }\label{subsection:optimizationResults}
The goal of the optimization is to design the phase-mask profiles to achieve the target unitary transformations from Eq.~\eqref{eq:2DUnitary}. The parameters of the MPLC are the ones listed in Subsection~\ref{subsection:Designparameters}. 

\begin{figure}[!htb]
    \centering
    \includegraphics[width=3.44 in ]{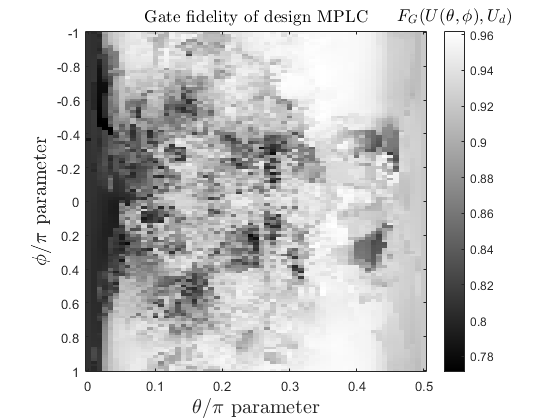}
    \caption{\label{fig:Fidelity_Design_Field_2} Predicted performance of the design MPLC system. We plot the theoretical gate fidelity $F_G\left(\bm{U}\left(\theta,\phi\right),\bm{U}_d\right)$ (obtained using Eq.~\eqref{eq:GateFide_FromFields}) between the target unitary $\bm{U}\left(\theta,\phi\right)$ transformation from Eq.~\eqref{eq:2DUnitary}, and the design one $\bm{U}_d$ obtained with the wavefront matching algorithm. The plot samples the full space of unitary transformations on a two-beam array.}
\end{figure}

We used the optimization algorithm to sample the whole space of unitary transformations from Eq.~\eqref{eq:2DUnitary}  with the following grid: $\theta \in [0, \pi/2]$ in steps of $\pi/120$, and $\phi \in [-\pi, \pi)$ in steps of $\pi/60$. Each design unitary $\bm{U}_d$ is the result of an optimization with $100$ iterations. Each optimization takes $\SI{53.806}{s}$ to run in a desktop computer with an eight core Intel i7-10700 processor at $\SI{2.9}{GHz}$ with $16$ GB of RAM (this same computer was used in the other data processing tasks).

The optimization results are shown in Fig.~\ref{fig:Fidelity_Design_Field_2}. Averaging $F_G$ over the $\bm{U}\left(\theta, \phi\right)$ space, we found the theoretical average gate fidelity to be $\bar{F}_G = 0{.}90 \pm 0.04$, where the uncertainty is the standard deviation. However, regions with significantly higher and lower fidelities than average can be distinguished. In particular, there is a strip region in $\theta \approx 0$ which gives the lowest fidelity, which is surprising since those transformations are close to the identity. The lower fidelity regions are mostly due to phase mismatch in the output states, which the wavefront matching algorithm was not able to correct even after increasing the number of iterations. We now describe the experimental implementation of unitary transformations in a two-beam array. Here we exploit the reconfigurability of the MPLC.

\section{Experiment}\label{sec:Experiment}

\begin{figure}[htb!]
    \centering
    \includegraphics[width = 4 in ]{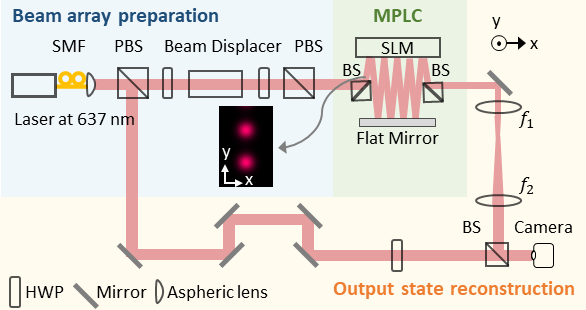}
    \caption{Experimental setup for the realization of an arbitrary unitary transformation on a two-beam array using a multi-plane light converter (MPLC). \textbf{Beam array preparation}. A polarizing beamsplitter (PBS) and a half wave plate (HWP) produce light linearly polarized at $45^{ \degree }$. A beam displacer shifts the spatial mode of photons with vertical polarization, and transmits spatially unshifted photons with horizontal polarization. The beam displacer creates the two-beam array. After the beam displacer, a HWP and a PBS select the input state. The transmitted light at the PBS is horizontally polarized, which is the working polarization of the spatial light modulator (SLM). \textbf{MPLC}. The MPLC is created with an SLM and a flat mirror. The beam array is inserted to the MPLC by a non-polarizing 50:50 beamsplitter. Five reflections are achieved on the SLM. After which, a second 50:50 beamsplitter is used to extract the output states from the MPLC. The efficiency of the five reflections on the MPLC (without considering the two beamsplitters) is $10 \%$. \textbf{Output state reconstruction}. A reference beam is created at the first PBS. The reference beam propagates through a delay line which compensates the optical path difference of the MPLC. A HWP at $45^{\degree}$ sets the reference beams's polarization to be horizontal. The output of the MPLC is imaged on a CMOS image sensor (camera) by a 4f-lens pair (magnification $1{.}25$). The imaged output states and the reference beam interfere after a beamsplitter and the interference pattern is recorded by the camera. Output state reconstruction is performed computationally from the interference pattern.}
    \label{fig:Exp_SetUp}
\end{figure}

The experimental setup is shown in Fig.~\ref{fig:Exp_SetUp}. Our light source is a laser diode with a wavelength of \SI{637.7 }{\nano\metre} with a FWHM bandwidth of \SI{1,3}{\nano\metre}  coupled to a single-mode fibre. We can describe the setup in three stages: beam array preparation, MPLC and output state reconstruction. The two-beam array is created with a birefringent beam displacer and polarization optics (see caption of Fig.~\ref{fig:Exp_SetUp} for details). We characterized the input beam array by measuring the beam size $\omega\left(z\right)$ at several propagation planes $z$. The measured beam characteristics were given in Subsection~\ref{subsection:Designparameters}, where they were used for defining the theoretical input beams. In the beam array preparation we have the ability to choose the input beam ($\ket{1}$ or $\ket{2}$), to the MPLC.

Our MPLC implementation uses five reflections on the SLM as the $P=5$ phase-mask planes. The SLM is a phase-only liquid crystal SLM (Hamamatsu X10468-07) with an effective area of $\SI{15.8}{m\metre}\times\SI{12}{m\metre}$ and a fill factor of $98 \%$. The SLM size is $792 \times 600$ pixels, with a pixel pitch of $\SI{20}{\micro\metre}$. The SLM is designed to have high throughput efficiency from $\SI{620}{}$ to  $\SI{1100}{\nano\metre}$. We measured a throughput efficiency of $63 \%$ at $\SI{637}{\nano\metre}$, and the manufacturer states a typical one of $82 \%$ at  $\SI{1064}{\nano\metre}$. Each pixel can be set to a value ranging from $0$ to $256$. We use the calibration provided by the manufacturer, which gives a linear response  from pixel value to phase and a pixel value of $118$ for a $2\pi$ modulation. We do not use the phase to correct aberrations provided by the manufacturer. The phase on the SLM $\Phi_{SLM}$ is obtained as follows  $\Phi_{SLM} = \frac{118}{2\pi } \left(\Phi \mod{ 2\pi} \right)$  where $\Phi$ is the phase at the five phase-mask planes directly obtained from the wavefront matching algorithm. 

We point out that we are not using the SLM in its most common mode of operation, which we now describe. Typically, when creating or manipulating optical modes with SLMs, the transformation is encoded on top of a modulation of a regular saw-tooth phase pattern, a blazed grating \cite{Davis:99,Bolduc:13}. This is a unitary process with each diffraction order carrying different information. Particularly, the light diffracted into the first-order contains the desired mode. Selecting a single diffraction order makes the process inherently non-unitary. This mode of operation of the SLM is used to achieve amplitude and phase modulation with a single hologram \cite{Davis:99,Bolduc:13}. It is also used as a selection mechanism due to the fact that SLM's diffraction efficiencies are lower than $100 \%$. In contrast, in the mode of operation we utilize, each phase-mask the SLM imparts is directly given by the wavefront matching algorithm. There is no underlying grating. While there may be technical sources of loss (e.g., fill-factor), our mode of operation can be fundamentally unitary.
 
While conducting  the experiment, we noticed the last two of the five design phase-mask profiles are the most important ones to realize the unitary transformations $\bm{U}\left(\theta, \phi\right)$. To that point, Table~\ref{table:ExperimentalGateFid_2vs5PP} shows the experimental gate fidelity between target transformations $\bm{U}\left(\theta,\phi = \pi/2\right)$ from Eq.~\eqref{eq:2DUnitary} and the experimentally obtained ones. The performance is same or better using solely the last two phase-mask profiles while setting the first three phase-mask planes to impart no phase. When using all five of the design phase-mask profiles, we observed that for many target unitary transformations the MPLC system produced a complicated scattered intensity distribution that was greatly different than the design output state. The physical mechanism for this behaviour is unknown and a subject for future investigation, however in the   \hyperref[sec:supplementarymaterial]{Supplementary material} we provide an example of how the performance of an MPLC greatly  decreases upon the presence of a phase-mask profile not taken into account in the design. For these reasons, the rest of the presented results use solely the last two of the five design phase-mask profiles. 

\begin{table}
\centering
\begin{tabular}{ c|c|c } 
 $\bm{U}\left(\theta,\phi = \pi/2\right)$    & $F_G$ five PPs           & $F_G$ two PPs \\
                  \hline
$\theta = \pi/2$    &    $0.91 \pm 0.01$        &    $0.914 \pm 0.009$          \\
\hline
$\theta = \pi/3$    &     $0.859 \pm 0.008$     & $0.85 \pm 0.05$             \\
\hline
$\theta = \pi/4$    &    $0.69 \pm 0.06$        &  $0.86 \pm 0.06$            \\
\hline
$\theta = \pi/5$    &   $0.83 \pm  0.02$        &   $0.85 \pm 0.06$           \\
\hline
$\theta = \pi/6$    &   $0.83 \pm 0.02$         &    $0.82 \pm 0.01$           \\
\end{tabular}
\caption{\label{table:ExperimentalGateFid_2vs5PP} Experimental gate fidelity  $F_G\left(\bm{U}\left(\theta,\phi = \pi/2\right), \bm{U}_e\right)$, calculated using Eq.~\eqref{eq:GateFide_FromFields}, between target unitary transformations $\bm{U}\left(\theta,\phi = \pi/2\right)$ from Eq.~\eqref{eq:2DUnitary} and the experimentally obtained ones $\bm{U}_e$ using all five or only the last two of the five phase-mask profiles (PP). The positions of the PPs were not changed when reducing the number of planes. We found using solely two PPs give same or better performance as five PPs. We indicate the standard deviation over five trials for statistical error.}
\end{table}

We exploit the reconfigurability of the SLM by experimentally sampling the two-dimensional unitary space $\bm{U}\left(\theta, \phi\right)$. The sampling used the following grid: $\theta \in [0, \pi/2]$ in steps of $\pi/120$ and $\phi \in [-\pi, \pi)$ in steps of $\pi/60$. The gate fidelity for each implemented transformation is obtained by using Eq.~\eqref{eq:GateFide_FromFields}. As for data acquisition, we set the phase-mask profiles on the SLM to perform a unitary $\bm{U}\left(\theta, \phi\right)$. For every input state $\ket{\psi} = \ket{m}$ ($m = 1, 2$), we record five images of the interference pattern. Each of these images are used to numerically reconstruct the optical field of the output state by off-axis holography (for a more detailed description of this method see for example \cite{picart2013digital,kreis2006handbook,digHolo}). Acquiring this set of ten images takes $\SI{25}{s}$. The off-axis holography routine takes $ \SI{44.12}{s}$ using MATLAB. Most of the time is taken by geometric transformations such as magnifying and flipping that account for the last 4f system in our experimental setup. This numerical  implementation can be optimized, see for example Ref.~\cite{digHolo}. The alignment procedure, and data acquisition are performed in LABVIEW. Further details of the alignment procedure can be found in the  \hyperref[sec:supplementarymaterial]{Supplementary material}.

\section{Results}\label{section:Results}
We first present the results of the method as described so far. We will then describe how the use of an  additional phase-mask plane improved the gate fidelity considerably.

\begin{figure}[htb!]
    \centering
    \includegraphics[width= 3.5 in]{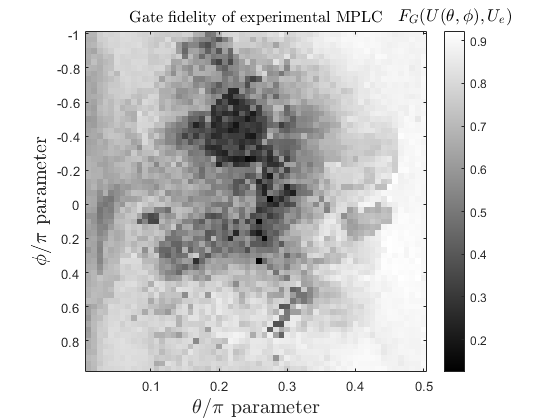}
    \caption{ Experimental gate fidelity $F_G\left(\bm{U}\left(\theta,\phi\right), \bm{U}_e\right)$ (calculated using Eq.~\eqref{eq:GateFide_FromFields}) between target unitary $\bm{U}\left(\theta,\phi\right)$ from Eq.~\eqref{eq:2DUnitary} and the unitary $\bm{U}_e$ experimentally implemented by the MPLC system. These results correspond to Part I: No correcting phase-mask. We sample the full space of the unitary transformations in a two-beam array given by Eq.~\eqref{eq:2DUnitary}. This plot is the experimental analogue of Fig.~\ref{fig:Fidelity_Design_Field_2}.}
    \label{fig:Exp_Gatefid_NoCPM}
\end{figure}

\subsection{Results - Part I: No correcting phase-mask}\label{sec:res_NoCPM}
The results for Part I--No correcting phase-mask are shown in Fig.~\ref{fig:Exp_Gatefid_NoCPM}. We show the experimental gate fidelity $F_G\left(\bm{U}_t, \bm{U}_e\right)$,  obtained by using Eq.~\eqref{eq:GateFide_FromFields}, between target unitary $\bm{U}_t$ and the experimentally implemented by the MPLC $\bm{U}_e$. Averaging over all the unitary space, we got an experimental average gate fidelity equal to $\bar{F}_g = 0{.}67 \pm  0{.}17$. This plot shows our experiment is able to implement a large set of unitary transformations with a gate fidelity greater  than $0.8$. However, there are regions of the unitary space implemented with an experimental gate fidelity lower than $0.5$ in the center of the plot. This contrast in gate fidelity is not expected from the design MPLC (Fig.~\ref{fig:Fidelity_Design_Field_2}), which predicts values of gate fidelity greater than $0.77$.  We next investigated ways to improve the MPLC in order to obtain a better performance over the full unitary space. 

We now describe the approach we followed for improving the average gate fidelity of our experiment. Observe that in Fig.~\ref{fig:Exp_Gatefid_NoCPM} there is a region at $\phi = \pi/2$ where the fidelity is higher than $0{.}8$ and independent of $\theta$. Such a strip of transformations accounts for the value of $\theta$ of the transformation $\bm{U}\left(\theta,\phi\right)$. We use an additional phase-mask plane to adjust the phase $\phi$ to create any target unitary $\bm{U}\left(\theta, \phi\right)$. This is the method experimentally used that we now report in the next subsection.

\begin{figure}[htb!]
    \centering
    \includegraphics[width= 3.44 in]{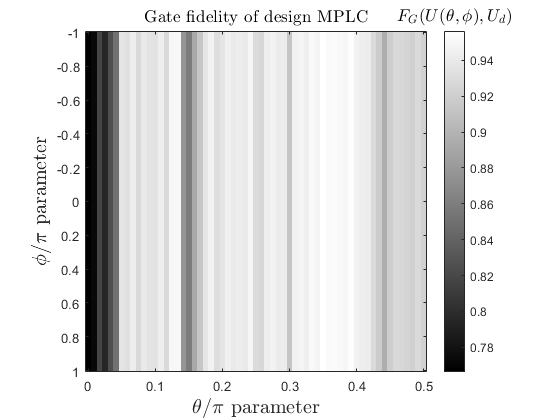}
    \caption{Performance of the design MPLC system plus a correcting phase-mask plane as described in Subsection~\ref{sec:res_CPM}. We plot the theoretical gate fidelity $F\left(\bm{U}\left(\theta,\phi\right),\bm{U}_d\right)$ (Eq.~\eqref{eq:GateFide_FromFields}) between the target unitary $\bm{U}\left(\theta, \phi\right)$ from Eq.~\eqref{eq:2DUnitary}, and the one implemented by the MPLC~$\bm{U}_d$. This MPLC uses the set of transformations $\bm{U}\left(\theta, \phi = \pi/2 \right)$ from Fig.~\ref{fig:Fidelity_Design_Field_2} and an additional phase-mask plane to sample the unitary transformations of Eq.~\eqref{eq:2DUnitary}.}
    \label{fig:Field_Fidelity_CorrectPhase}
\end{figure}

\subsection{Results - Part II: With correcting phase-mask}\label{sec:res_CPM}
Our experimental results presented in Fig.~\ref{fig:Exp_Gatefid_NoCPM} show a non-ideal performance of our MPLC over the unitary space. In other words, there are regions  with contrasting fidelities. To improve the performance of the MPLC, we use the line of designs along $\bm{U}\left( \theta, \phi = \pi/2 \right)$ and an additional phase-mask profile located at a sixth phase-mask plane of the MPLC to achieve any other transformation $\bm{U}\left( \theta, \phi \right)$. This phase-mask applies a phase difference 
\begin{equation}
\Delta\phi = \left(\phi-\pi/2\right)/2, \label{eq:CPM_DeltaPhi}   
\end{equation}
to the top beam and $-\Delta\phi$ to the bottom one. To test the correcting phase-mask approach,  we applied it to the design MPLC described in Section~\ref{subsection:optimizationResults}. The resulting theoretical gate fidelity of the design MPLC with correcting phase-mask is shown in Fig.~\ref{fig:Field_Fidelity_CorrectPhase}. It has an average gate fidelity $\bar{F}_G = 0{.}92 \pm 0{.}04$, which has a percentage difference of $2.2 \%$ respect to the one of Fig.~\ref{fig:Fidelity_Design_Field_2}. Thus our correcting phase-mask approach preserves the performance of the original theoretical system design. The lower performance of transformations with $\theta = 0 - 6\pi/120$ is due to a phase gradient of $\pi$ across each beam, which is not fixed by the wavefront matching algorithm.  

\begin{figure}[htb!]
    \centering
    \includegraphics[width= 3.44 in]{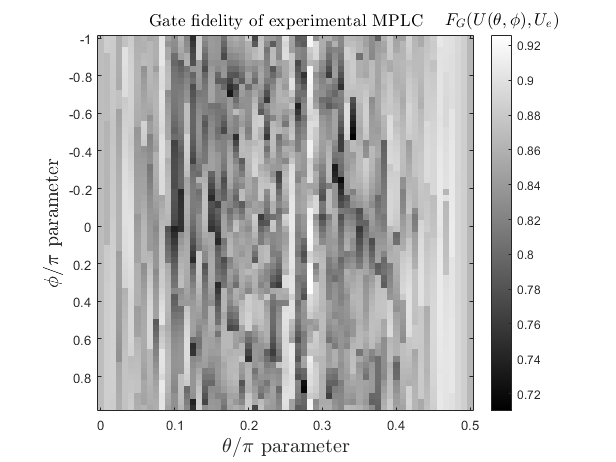}
    \caption{Experimental gate fidelity $F_G \left( \bm{U}\left(\theta,\phi\right), \bm{U}_e\right)$ (Eq.~\eqref{eq:GateFide_FromFields}) between target unitary $\bm{U}\left(\theta,\phi\right)$ from Eq.~\eqref{eq:2DUnitary} and the unitary $\bm{U}_e$ experimentally implemented. These results correspond to Part II: With correcting phase-mask.}
    \label{fig:GateFid_CPM}
\end{figure}

For the experimental implementation of the correcting mask approach, we used an additional SLM located at the output plane of the MPLC system from Fig.~\ref{fig:Exp_SetUp}. This SLM is a Pluto-2 phase only SLM with a dielectric mirror at its back plane to enhance reflectivity in the range of $\SI{730}{}-\SI{950}{\nano\metre}$. It achieves a maximum modulation of $2{.}4\pi$ with a pixel value of $\SI{255}{}$ at $\SI{637}{\nano\metre}$. Implementing the correcting mask approach yielded the results shown in Fig.~\ref{fig:GateFid_CPM}. Averaging over the unitary transformation space, we obtained an experimental average gate fidelity of $\bar{F}_G = 0{.}85 \pm 0{.}03$. This is a much improved result when compared to Fig.~\ref{fig:Exp_Gatefid_NoCPM} (note the change in scale).

\begin{figure}[htb!]
    \centering
    \includegraphics[width= 3.5 in]{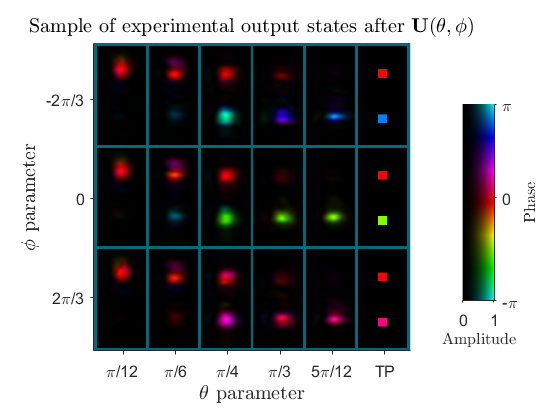}
    \caption{Field of experimentally obtained output states for different transformations $\bm{U}\left(\theta,\phi\right)$. The blue grid separates different states corresponding to a different transformation. The values of $\theta$ and $\phi$ are indicated on the axes. In this sample, the input state is the beam $\ket{1}$ of the beam array. The last column shows the target phase (TP) for the fields on the same row. The field amplitude of each state was normalized to have the same maximum amplitude. This was done for image visibility purposes.}
    \label{fig:Exp_fields}
\end{figure}

Fig.~\ref{fig:Exp_fields} shows a sample of experimentally obtained output states for different unitary transformations $\bm{U}\left(\theta,\phi\right)$ when the input state is the beam $\ket{1}$. The amplitude and phase of the fields are shown in the brightness and hue of the plot respectively. This figure explicitly demonstrates the ability of an MPLC to implement the full gamut of unitary transformations. The obtained output states cover the full range of amplitude and phase a state in the beam array can have. As a result, we are also able to generate any state in the two-beam array. 

\section{Discussion}\label{sec:discussion}
As we mentioned earlier, the modeled performance of the design MLPC (Fig.~\ref{fig:Fidelity_Design_Field_2}) is significantly higher than the observed experimental performance (Fig.~\ref{fig:Exp_Gatefid_NoCPM}). We now outline an example of potential experimental imperfections that effectively decrease the performance of an MPLC system. Examples of such experimental imperfections include light scattering, misalignment of the system,  mismatch between design and experimentally available beams, positioning of phase-mask planes, SLM's surface curvature, inherent grid in pixel structure of an SLM, etc. Modelling these imperfections is challenging and they become more prominent with an increasing number of phase-mask planes.  We calculated the  effect of a not unrealistic phase-mask distortion in an MPLC. Such a phase-mask perturbation has an average (averaged over the five locations of the reflections on the SLM) gradient equal to $0.058 \pm 0.02$~ $\SI{ }{rad/\text{SLM pixel}}$. This perturbation causes a $20\%$ drop in the theoretical gate fidelity compared to the original design MPLC. The details of this calculation can be found in the  \hyperref[sec:supplementarymaterial]{Supplementary material}. 

An MPLC system could be used as an auxiliary device in a quantum computer e.g., for state preparation or in the implementation of gates. For such an application an SLM with higher efficiency is needed. At present there are SLMs with less than $3\%$ optical loss within a  narrow band light spectral region. This low loss is achieved by using a dielectric mirror behind the liquid crystal array. 

The realization of reconfigurable unitary transformations in a high-dimensional beam array is the ultimate goal of an MPLC system. Such a device would allow one to use linear optics in free-space for implementing and studying boson sampling experiments, the implementation of quantum gates and random walks which have many applications, see for example \cite{venegas2012quantum}. Another interesting application is the 
use of an MPLC as a \textit{Meta-optic} device, implementing arbitrary spatial transformations such as imaging, spatial filtering, or as a \textit{spaceplate} \cite{reshef2021optic}.

\section{Conclusions}
As suggested by the experimentally obtained results, an MPLC system is a challenging system to characterize and it is sensitive to experimental errors which are hard to eliminate. On the other hand, the reconfigurability feature of an MPLC allows experimentalists to adjust input states and phase-mask profiles of the MPLC system, and to measure the output states after such changes. Such an experimental information can potentially be used to optimize the phase-mask profiles of the MPLC and account for  experimental imperfections that were not taken into account at the design stage. There are different approaches to this experimentally optimized version of an MPLC system. For example the phase-mask profiles could be optimized by machine learning techniques e.g., neural networks, genetic algorithms, stochastic sampling or by the optimization of an analytical model for the phase-mask profiles. This is undergoing work in our lab and we leave details for future work.   

In summary, we experimentally demonstrated a reconfigurable MPLC capable of implementing arbitrary unitary transformations in a two-beam array. To do so, we had to substantially modify the design created by the standard method, wavefront matching optimization. Firstly, we had to omit all but the last two of the five phase profiles output from the optimization. And secondly, we had to use a second SLM to apply an additional phase profile after the nominal MPLC. The latter phase profile was determined from the characterization of that MPLC. Nonetheless, we demonstrate a gate fidelity of $\bar{F}_G = 0{.}85 \pm 0{.}03$ that for the first time is found by testing and averaging over all unitary transformations. Thus, this work demonstrates the usefulness of MPLC systems and how to use them in a reconfigurable way, leading to a novel photonics tool that can benefit both classical and quantum optics.

\section*{Acknowledgments}
This work was supported by the Canada Research Chairs Program, the Natural Sciences and Engineering Research Council, the Canada First Research Excellence Fund (Transformative Quantum Technologies), the U.S. Department of Energy QuantISED award, and the Brookhaven National Laboratory LDRD Grant No. 22-22.

\bibliography{RecUnitaryTsOpticalBeamArrays}
\bibliographystyle{unsrt}

\newpage
\section*{Supplementary material}\label{sec:supplementarymaterial}

\subsection*{Number of reflections in an MPLC system}
In this section we calculate the number of reflections achievable in an MPLC composed by a flat mirror and an SLM as shown in Fig.~\ref{fig:MPLC_BeamArray_PMs}. The SLM-to-mirror distance is $L$ and the insertion angle is~$\tau$. The optical path length between two consecutive reflections is $2L\sec\tau.$ We assume a Gaussian beam with beam waist $\omega_0$ at the first plane. The beam size is $\omega\left(z\right) = \omega_0\left(1+\left(z/z_0\right)^2\right)^{1/2}$ where $z$ is the propagation distance. Thus, the beam size at reflection $p$ on the SLM is given by 
\begin{equation}
\omega_0\left(1+\left(2Lp\sec\tau/z_0\right)^2\right)^{1/2}. \label{eq:beamsize_pPlane}    
\end{equation}
Now we count the number of reflections on the SLM, such a number is constrained by the SLM's dimensions. The other constraint is avoiding overlap between consecutive reflections. Given that $99\%$ of a Gaussian beam power lies within a circle of radius $1{.}5\omega\left(z\right)$ \cite{saleh2019fundamentals}. We consider there is no overlap if there is at least a distance equal to $1{.}5\omega\left(z\right)$ (plus a few pixels to distinguish the reflections) between two reflections. 

\begin{figure*}[htb!]
\centering
Reflections on SLM in an MPLC \\
      \hspace{2.5 mm} a)~ $L = \SI{1}{\centi\metre}$  \hspace{47 mm} b)~ $L = \SI{10}{\centi\metre}$   \\
     \includegraphics[width= 2.6 in]{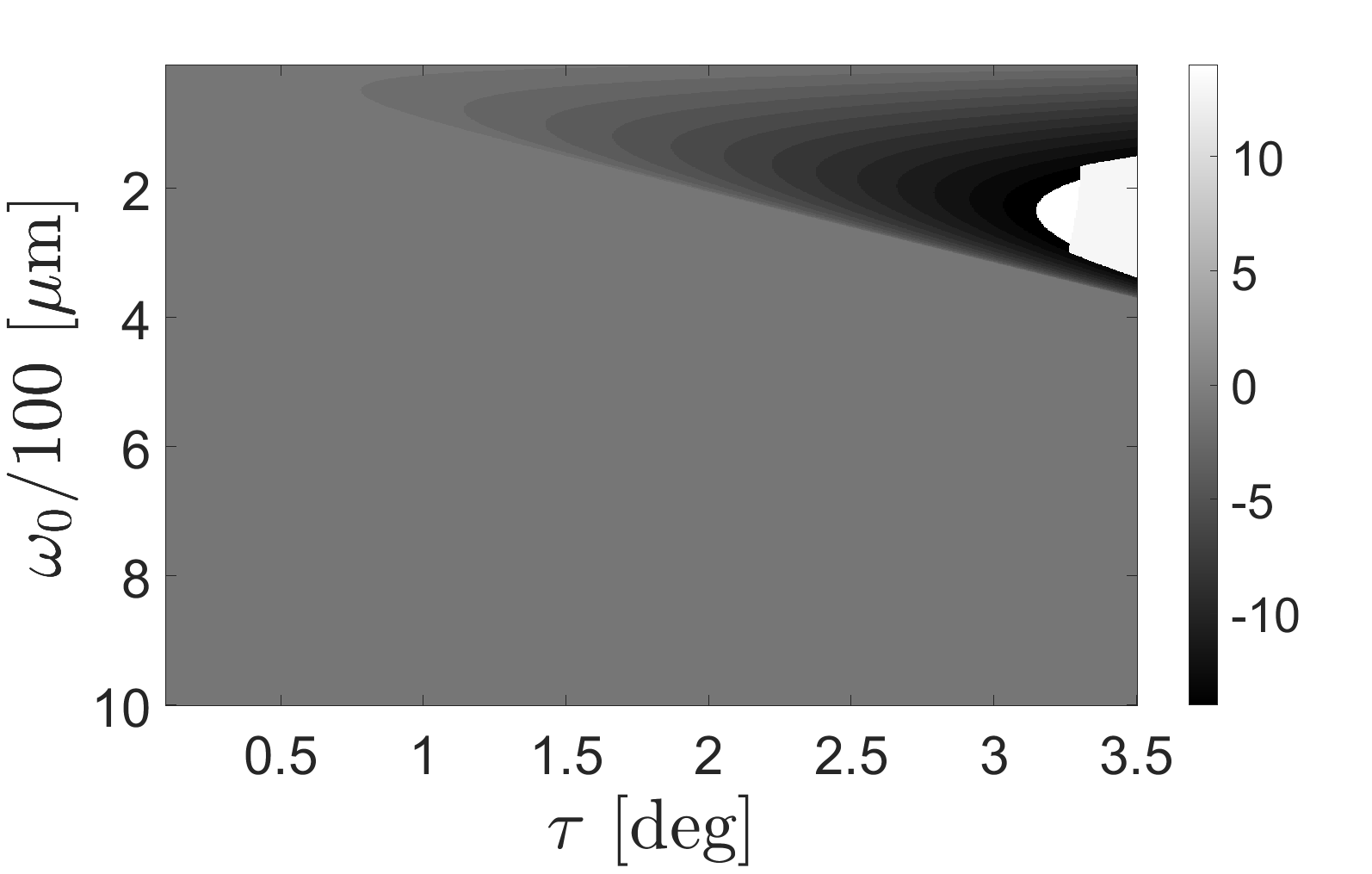}
      \includegraphics[width= 2.6 in]{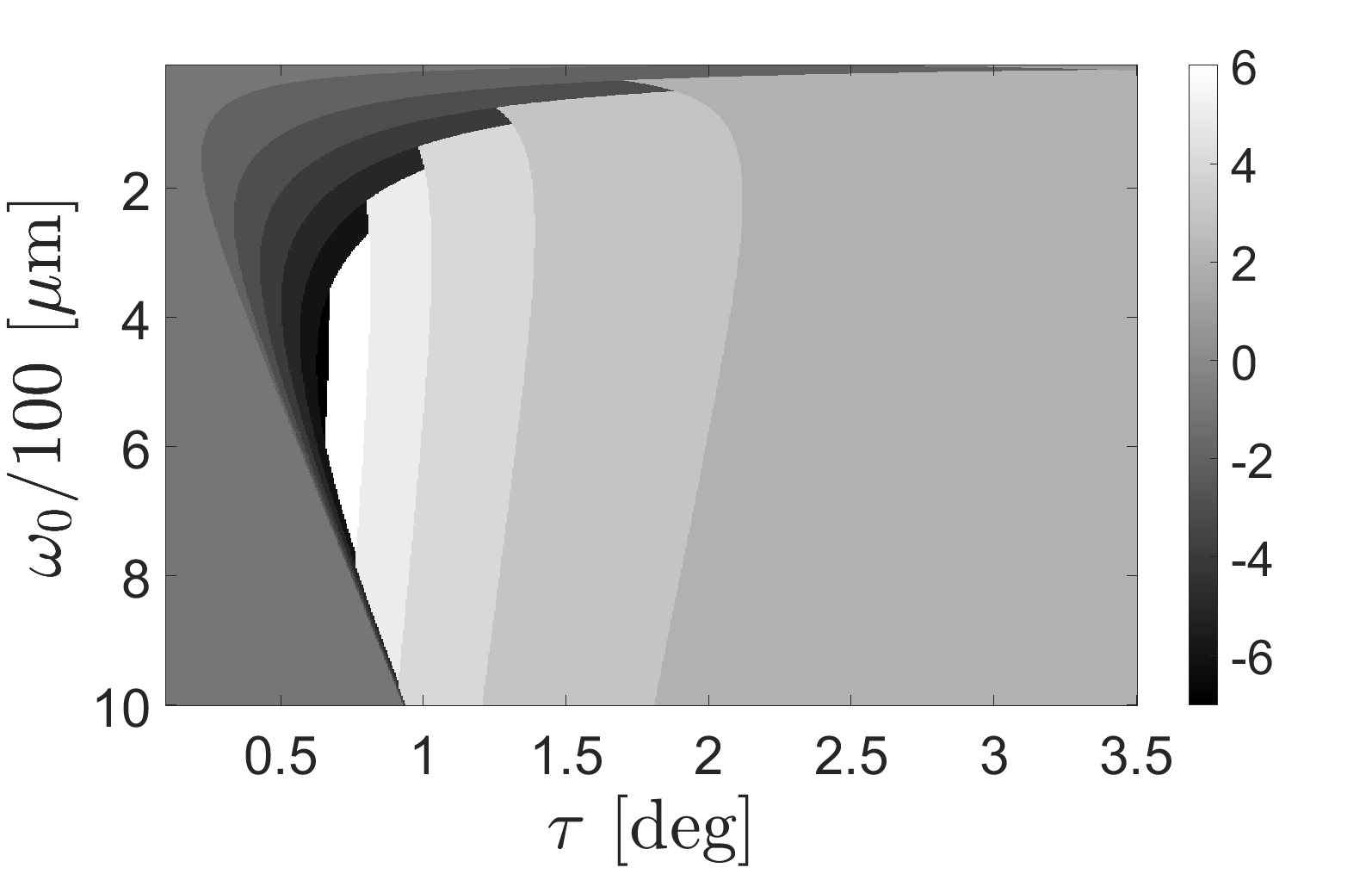}
  \caption{Number of reflections on an SLM (dimensions given in Section \ref{sec:Experiment}) in an MPLC system as a function of the insertion angle $\tau$ and the beam waist $\omega_0$ located at the first reflection plane. In a)~the SLM-mirror distance $L$ is set to $ \SI{1}{cm}$ and $\SI{10}{cm}$ in b). The regions with negative values indicate overlap between reflections at that particular reflection. For $L = \SI{10}{\centi\metre}$ the angle $\tau$ is restricted to be greater than $\SI{0.7}{\degree}$ to achieve multiple reflections. As $L$ decreases, the beam size is constrained to a few hundreds of microns.}    \label{fig:ReflectionsDesign_ParameterSpace}
\end{figure*}

\begin{figure*}[htb]
    \centering
    \hspace{-1 mm} Maximum number of phase-mask planes in an MPLC  \\
    \includegraphics[width = 4.5 in]{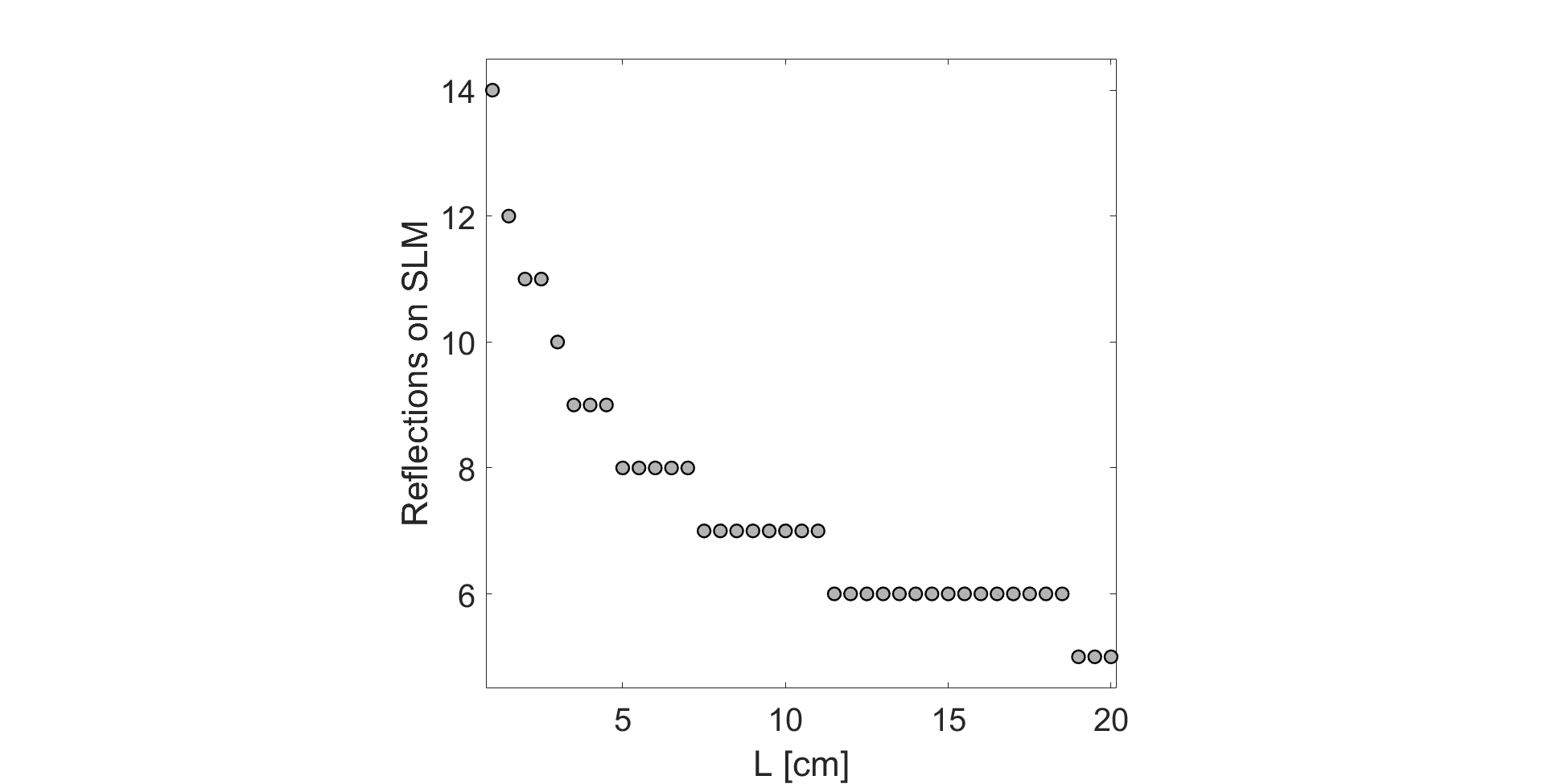}
  \caption{Maximum number of phase-mask planes in an MPLC system (built with a single SLM) as a function of the mirror-SLM separation $L$.}    \label{fig:ReflectionsvsL}
\end{figure*}

We search the number of reflections numerically using Eq.~\eqref{eq:beamsize_pPlane} and above mentioned constraints. Fig.~\ref{fig:ReflectionsDesign_ParameterSpace} shows the number of reflections on the SLM (SLM details given in Section~\ref{sec:Experiment}) as a function of $\omega_0$ and $\tau$ for two different distances $L$. The maximum number of reflections for different values of $L$ is shown in Fig.~\ref{fig:ReflectionsvsL}. More reflections are achieved as the value of $L$ decreases, as expected. As $L$ decreases, the amount of diffraction also decreases. We anticipate this would have an effect in the phase-mask profiles required to make a target unitary transformation.  

\subsection*{Alignment of the MPLC}
In what follows we give further information of the experimental implementation of our MPLC system. The choice of parameters of the MPLC and beam array characterization was detailed in Subsection~\ref{subsection:Designparameters}. The experimental setup was described in Section~\ref{sec:Experiment}. We will now explain how to set the position of the phase-mask planes on the SLM.  

We use a `diffractive knife-edge' method to  locate the centers of the beam reflections on the SLM. This method is a diffraction analogue of the  widely known `knife-edge method' for Gaussian beam characterization \cite{Khosrofian:83}. The knife-edge method records the transmitted light power after a  knife-edge is moving transversal to a beam. Such information is used to obtain the beam center and  beam size \cite{Khosrofian:83} at the knife plane. We now describe how this technique can be emulated with an SLM. A phase-mask of random phase values (a random phase-mask)  diffracts light in all directions, which effectively creates loss in the far field. Thus, a phase-mask with a translating rectangular pattern of random phase values has the same effect as the translating knife-edge. In an MPLC, the random phase-mask should be chosen so that it covers the full beam without overlap between different reflections on the SLM. The transmitted power after passing by each reflection is used for obtaining the center and beam size at each phase-mask plane following the same computation from the knife-edge method (see for example Ref.~\cite{Khosrofian:83}). This computation requires post-processing as the data needs to be noise filtered to be used for computing a derivative.

A rough estimate of the beam size and beam center can be done with a modification of the diffractive knife-edge method. One can use a random phase-mask of the expected beam size and scan it through the SLM.  When the random phase-mask overlaps with a beam reflection, the beam will disappear in the far-field. Position and size of the random phase-mask can be adjusted for an estimate of the beam size and center. 

In the alignment of our experiment, described in Section~\ref{sec:Experiment}, we used a random phase-mask obtained from a Gaussian distribution with center in 60 and standard deviation of ten. Each value of the random phase-mask is converted to an unsigned integer to be displayed on the SLM.   

For an MPLC, the centers of the phase-mask profiles need to be further optimized. We use a trial transformation $\bm{U}\left(\theta, \phi\right)$ (Eq.~\eqref{eq:2DUnitary}) and use the intensity on the camera after we apply a new single phase-mask at a time.  The figure of merit to optimize is the `intensity fidelity' $F_I\left(I_d,I_e\right)$ defined by the following equation
\begin{equation}
    F_I\left(I_d,I_e\right) = \int I_d\left(y\right)I_e\left(y\right) dy,
\end{equation}
where $I_d\left(y\right) $ is a theoretical intensity and $I_e\left(y\right)$  is the experimental intensity we are optimizing. Such intensities $I_y\left(y\right)$ along $y$ can be obtained from an x-y intensity $I\left(x,y\right)$ by integrating along $x$, i.e., $I_y\left(y\right) = \int I_y\left(x,y\right)dx/\left( \int  \left( \int I(x,y) dx \right)^2 dy \right)^{1/2}$. This intensity fidelity is a parameter between zero (not matching) and one (perfect matching). 

We start by the first phase-mask plane, the center of the first phase-mask is scanned along the $y$ direction in steps of one pixel while the other phase-masks are set to zero. The experimental intensity $I_{e}\left(x,y\right)$ is measured on the CCD camera at the last plane of the MPLC system. This process was previously simulated with the design phase-mask profiles giving us a theoretical design intensity $I_d\left(x,y\right)$. The center of the first phase-mask profile is scanned to maximize the intensity fidelity $F_I\left(I_d,I_e\right)$. We repeat the same procedure for the $x$ direction. We noticed the $y$ direction is more sensitive to changes compared to the $x$ direction.

Having optimized the first phase-mask, we apply the first two phase-mask profiles and repeat the fine tuning procedure only for the second phase-mask plane. We repeat this procedure with the rest of the phase-mask planes. Ideally the found positions are fixed and should work for any transformation~$\bm{U}\left(\theta, \phi\right)$. 

\begin{figure}[htb!]
    \centering
    \includegraphics[width= 3.3 in]{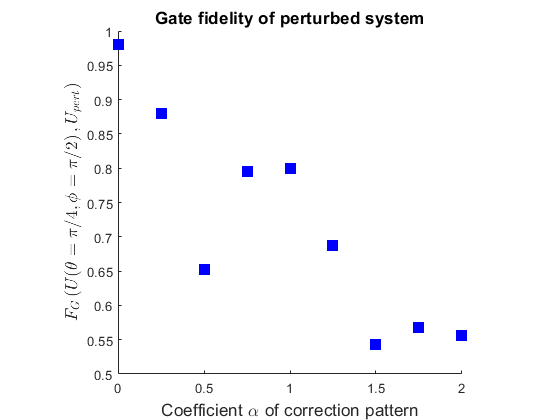}
    \caption{Theoretical gate fidelity $F_G\left(\bm{U}\left(\theta = \pi/4,\phi = \pi/2 \right), \bm{U}_{pert}\right)$ between target unitary $\bm{U}\left(\theta = \pi/4,\phi = \pi/2 \right)$ and the unitary $\bm{U}_{pert}$ implemented by a perturbed MPLC system. The phase-mask perturbation is given by $\alpha\Phi_{pert}$ with $\Phi_{pert}$ being the correction phase pattern provided by the manufacturer of the SLM. This phase was used as an example of a not unrealistic perturbation that can appear in an MPLC system.}
    \label{fig:GateFid_Mismatch}
\end{figure}

\subsection*{ Effect of a perturbation on the phase-mask planes}
We found a mismatch in the performance of the design MPLC (Fig.~\ref{fig:Fidelity_Design_Field_2}) and the experimental one (Fig.~\ref{fig:Exp_Gatefid_NoCPM}). We now give an example of how a perturbation to the imparted phase-mask profiles decreases the performance of an MPLC. 

\begin{figure*}[htb]
    \centering
        Phase-mask profiles for the perturbed MPLC with coefficient $\alpha = 0.25$\\
        \includegraphics[width = 5.1 in]{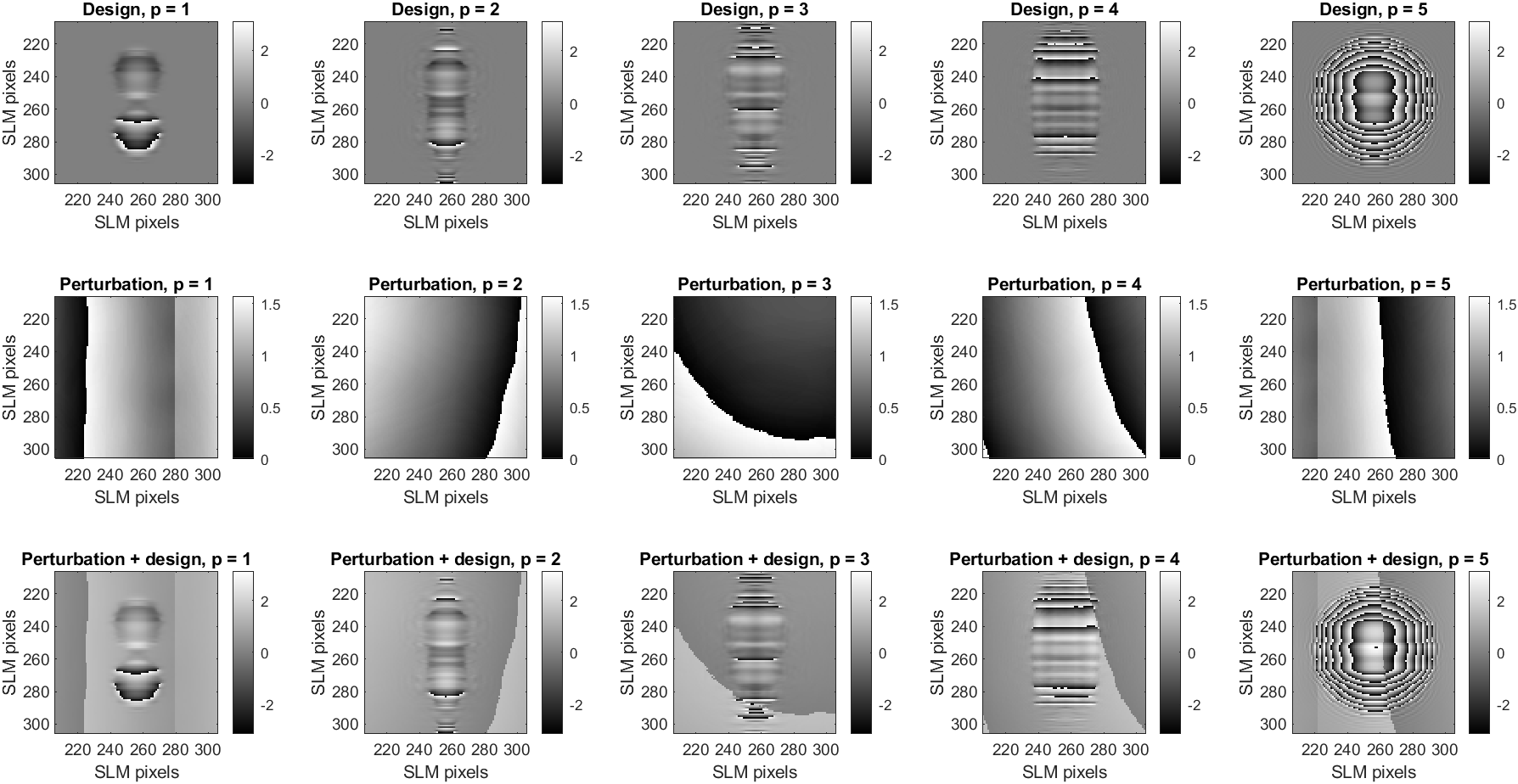 }
  \caption{ Phase-mask profiles for the design and perturbed MPLC systems. First row shows the design phase-mask profiles at each of the $p = 1,\ldots,5$ phase-mask planes for the unitary $\bm{U}\left(\theta = \pi/4, \phi = \pi/2 \right)$. Second row shows the perturbation phase-mask   $\alpha\Phi_{pert}$ with $\alpha = 0.25$. The third row shows the phase addition of rows one and two according to Eq.~\eqref{eq:phase_perturbed_MPLC}.}    \label{fig:PhasePlanes_alpha_0_25}
\end{figure*}

The procedure is as follows. We used the design phase-mask profiles $\Phi\left(x,y\right)$ obtained in Subsection~\ref{subsection:optimizationResults} to implement  the unitary transformations $\bm{U} \left( \theta, \phi \right)$ (from Eq.~\eqref{eq:2DUnitary}). We added a phase-mask perturbation  $\alpha\Phi_{pert}\left(x,y\right)$. Here,  $\alpha$ is a coefficient and $\Phi_{pert}\left(x,y\right)$ is the phase perturbation. The phase-mask profile imparted by the perturbed MPLC system is given by the sum of  $\Phi\left(x,y\right)$ and  $\alpha\Phi_{pert}$ modulo $2\pi$, which is mathematically written as follows 
\begin{equation}
 \Phi\left(x,y\right) + \alpha\Phi_{pert}\left(x,y\right) 
    \mod{ 2\pi}. \label{eq:phase_perturbed_MPLC}
\end{equation}
We took $\Phi_{pert}$ to be the one provided by the manufacturer to correct for the surface curvature of the SLM, i.e., the SLM applies an spatially varying phase even when no phase-mask is being displayed on it; the correction pattern $\Phi_{pert}$ is meant to cancel such an effect. Such a phase perturbation has an average gradient equal to $0.058 \pm 0.02 ~ \SI{ }{rad/\text{SLM pixel}}$. The average was taken over the five reflections on the SLM. 

We now assess the performance of the perturbed MPLC system. We propagate each of the beams from the beam array ($\ket{1}$ or $\ket{2}$) through the perturbed MPLC and use Eq.~\eqref{eq:GateFide_FromFields} to obtain a gate fidelity. Fig.~\ref{fig:GateFid_Mismatch} shows the theoretical gate fidelity between the transformation implemented by the perturbed MPLC (with varying coefficient $\alpha$) and a target unitary $\bm{U}\left(\theta = \pi/4,\phi = \pi/2 \right)$. The gate fidelity dramatically drops with increasing values of $\alpha$. This trend also appears for transformations with different values of $\theta$. 

\begin{figure*}[htb]
    \centering
    Phase-mask profiles for the perturbed MPLC with coefficient $\alpha = 1.25$\\
    \includegraphics[width = 5.1 in]{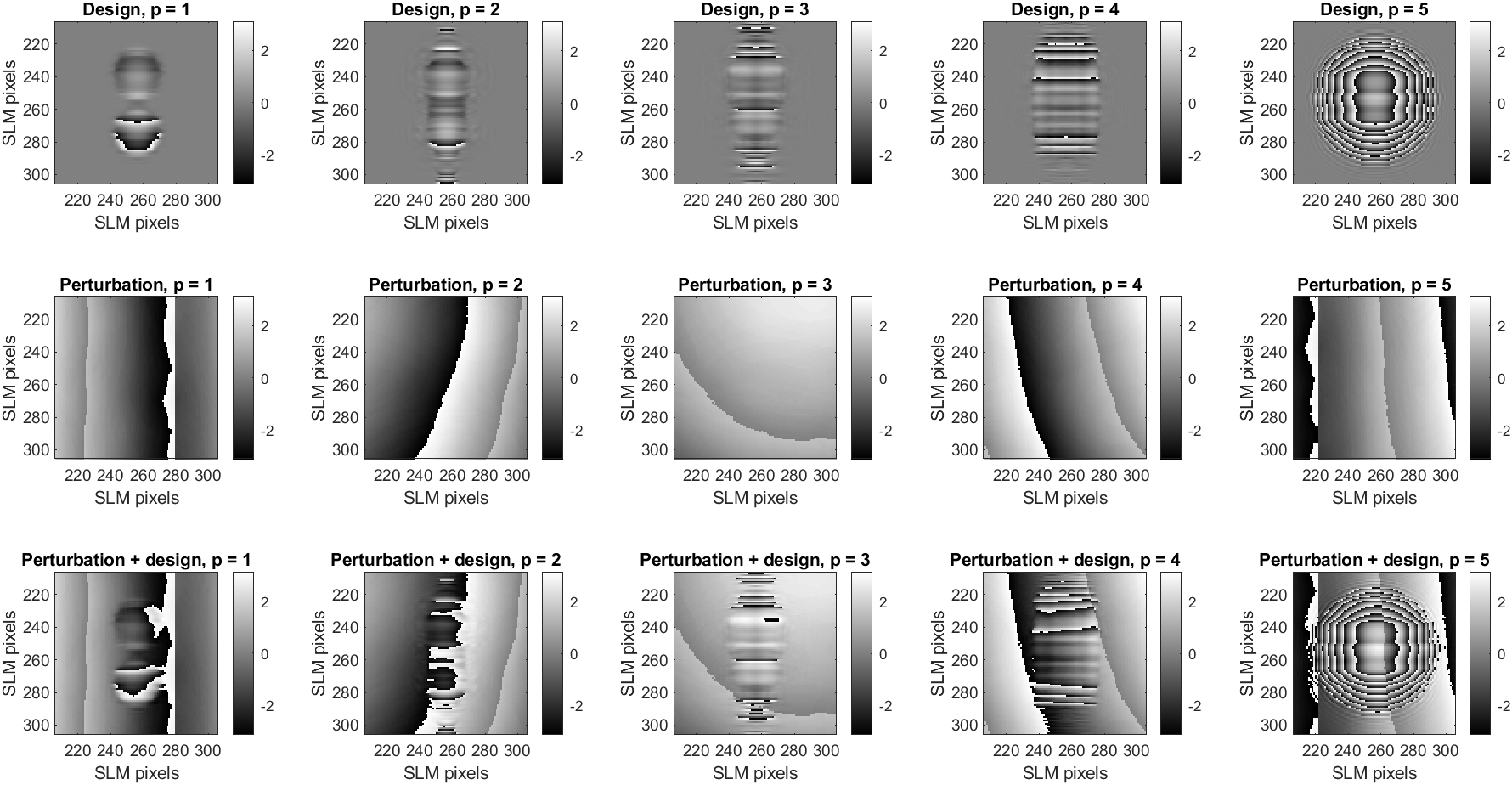 }
  \caption{ Phase-mask profiles for the design and perturbed MPLC systems. First row shows the design phase-mask profiles at each of the $p = 1,\ldots,5$ phase-mask planes for the unitary $\bm{U}\left(\theta = \pi/4, \phi = \pi/2 \right)$. Second row shows the perturbation phase-mask   $\alpha\Phi_{pert}$ with $\alpha = 1.25$. The third row shows the phase addition of rows one and two according to Eq.~\eqref{eq:phase_perturbed_MPLC}.}    \label{fig:PhasePlanes_alpha_1_25}
\end{figure*}

The phase-mask profiles obtained from this example are shown in Figs.~\ref{fig:PhasePlanes_alpha_0_25}~and~\ref{fig:PhasePlanes_alpha_1_25}  for $\alpha = 0.25$ and $\alpha = 1.25$ respectively. This example shows that not unrealistic phase perturbations dramatically affect the performance of an MPLC system. We finally mention the correction pattern $\Phi_{pert}$ was not used in our experiment because it distorted the beams even when no other phase was added to $\Phi_{pert}$.  

\end{document}